\documentclass{aa}

\usepackage{graphicx}
\usepackage{txfonts}
\usepackage{booktabs}
\usepackage{multirow}
\usepackage{hyperref}
\usepackage{siunitx}

\newcommand{\feh}{[Fe/H]}
\newcommand{\ali}{A(Li)}

\newcommand{\eccentricity}{$\langle e \rangle$}
\newcommand{\zmax}{$\langle Z_{\rm{max}} \rangle$}
\newcommand{\lz}{$\langle L_z \rangle$}
\newcommand{\mass}{$\langle M \rangle $}
\newcommand{\age}{$\overline{t}_{\star}$}
\newcommand{\teff}{$T_{\rm eff}$}

\begin{document} 

    \title{Probing the origins}
    
    \subtitle{II. Unravelling lithium depletion and stellar motion: Intrinsic stellar properties drive depletion, not kinematics}

    \titlerunning{Origins II -- Intrinsic stellar properties drive lithium depletion, not kinematics}

    \author{M.~L.~L.~Dantas\inst{\ref{ia_puc}, \ref{astroing_puc}}
           \and
           R.~Smiljanic\inst{\ref{camk}}
           \and
           D.~Romano\inst{\ref{inaf_bologna}}
           \and
           G.~Guiglion \inst{\ref{uheidelberg}, \ref{mpa}, \ref{aip}}
           \and
           L.~Magrini\inst{\ref{inaf_oaa}}
           \and
           P.~B.~Tissera\inst{\ref{ia_puc}, \ref{astroing_puc}}
           \and
           R.~S.~de Souza\inst{\ref{herts}, \ref{iag}, \ref{unc}}
           }

    \institute{
    Instituto de Astrofísica, Pontificia Universidad Católica de Chile, Av. Vicuña Mackenna 4860, Santiago, Chile \label{ia_puc}
    \and
    Centro de Astro-Ingeniería, Pontificia Universidad Católica de Chile, Av. Vicuña Mackenna 4860, Santiago, Chile \label{astroing_puc}\\
    \email{mlldantas@protonmail.com}
    \and
    Nicolaus Copernicus Astronomical Center, Polish Academy of Sciences, ul. Bartycka 18, 00-716, Warsaw, Poland \label{camk}
    \and
    INAF -- Osservatorio di Astrofisica e Scienza dello Spazio, Via Gobetti 93/3, 40129 Bologna, Italy \label{inaf_bologna}
    \and
    Zentrum f\"ur Astronomie der Universit\"at Heidelberg, Landessternwarte, K\"onigstuhl 12, 69117 Heidelberg, Germany \label{uheidelberg}
    \and
    Max Planck Institute for Astronomy, K\"onigstuhl 17, 69117, Heidelberg, Germany \label{mpa}
    \and
    Leibniz-Institut f{\"u}r Astrophysik Potsdam (AIP), An der Sternwarte 16, 14482 Potsdam, Germany \label{aip}
    \and
    INAF -- Osservatorio Astrofisico di Arcetri, Largo E. Fermi, 5, 50125 Firenze, Italy \label{inaf_oaa}
    \and
    Centre for Astrophysics Research, University of Hertfordshire, College Lane, Hatfield, AL10~9AB, UK \label{herts}
    \and 
    Instituto de Astronomia, Geofísica e Ciências Atmosféricas, Universidade de São Paulo, Rua do Matão 1226, 05508-090, São Paulo, Brazil \label{iag}
    \and 
    Department of Physics \& Astronomy, University of North Carolina at Chapel Hill, NC 27599-3255, USA \label{unc}
    }
   
    \date{Received XXX; accepted XXX}

    % \abstract{}{}{}{}{} 
    % 5 {} token are mandatory
 
\abstract
% context heading (optional)
{Lithium (Li) is a complex yet fragile element, with many production pathways but is easily destroyed in stars. Previous studies observe that the top envelope of the distribution of Li abundances A(Li) in super-solar metallicity dwarf stars shows signs of Li depletion, contrary to expectations. This depletion is thought to result from the interplay between stellar evolution and radial migration.}
% aims heading (mandatory)
{In Paper I, we classified a stellar sample from the thin disc with a broad range in metallicity as being churned outwards or inwards, or as stars where angular momentum was preserved (a category including blurred and undisturbed stars, which our method does not separate). In this paper (Paper II), we delve deeper by analysing our entire metallicity-stratified sample along with their dynamic properties, focusing on the connection between radial migration and Li depletion.}
% methods heading (mandatory)
{We analysed the chemo-dynamics of a set of 1188 thin-disc dwarf stars observed by the \textit{Gaia}-ESO survey, previously classified into six metallicity-stratified groups via hierarchical clustering (HC), ranging from metal-poor to super-metal-rich. We examined several features, such as effective temperatures, masses, and dynamic properties. We also implemented a parametric survival analysis using penalised splines (logistic distribution) to quantify how stellar properties and motion (or migration) direction jointly influence Li depletion patterns.}
% results heading (mandatory)
{Stars in our sample that seemingly churned outwards are predominantly Li-depleted, regardless of their metallicities. These stars are also the oldest, coldest, and least massive compared to those in the same HC group that either churned inwards or kept their orbital radii. Our survival analysis confirms temperature as the primary driver of Li depletion, followed by metallicity and age, while migration direction shows negligible influence. Additionally, the proportion of outward-churned stars increases with increasing metallicity, making up more than 90\% of our sample in the most metal-rich group.}
% conclusions heading (optional)
{The increasing proportion of outward-churned stars with higher metallicity (and older ages) indicates their dominant influence on the overall trend observed in the [Fe/H]-A(Li) space for stellar groups with [Fe/H]>0. The survival model reinforces the finding that the observed Li depletion stems primarily from intrinsic stellar properties (cool temperatures, higher metallicity, and old ages) rather than migration history. This suggests the metallicity-dependent depletion pattern emerges through stellar evolution rather than Galactic dynamical processes.}


   \maketitle
   
%-------------------------------------------------------------------
\section{Introduction} \label{sec:intro}

Among all the elements in the Universe, lithium (Li) stands out as one of the most intriguing. It is the third most abundant element produced during the Big Bang nucleosynthesis, and it is also formed shortly thereafter through the decay of beryllium ($^7$Be) into $^7$Li \citep[e.g.][]{Pinsonneault1997, Fields2020}. Lithium has diverse production pathways, including nuclear burning in stellar interiors \citep[e.g. in Li-rich giants; see for instance][]{SackmannBoothroyd1992, Sayeed2024}, novae explosions \citep[][see also \citealt{Molaro2016, Rukeya2017, Borisov2024}, and references therein]{Starrfield1978}, and cosmic-ray spallation \citep[e.g.][]{OliveSchramm1992, Knauth2000}. However, Li is a delicate element, making its abundance highly variable. It undergoes significant changes and destruction due to stellar evolutionary processes \citep[e.g.][]{Smiljanic2020, Randich2021}. For instance, the Sun itself has experienced massive Li depletion, losing more than two orders of magnitude in its abundance from the initial value, as recorded in meteorites \citep{Lodders2009}, since the genesis of the Solar System \citep[see e.g.][for the current Li abundance]{Asplund2009, Wang2021}.

This complex interplay of production and destruction mechanisms means Li abundances A(Li)s must be interpreted cautiously. The extent of Li depletion is strongly metallicity-dependent: for metal-poor stars (\feh $\lesssim -1$), the canonical depletion threshold lies near \teff $\simeq 5700$ K \citep{Rebolo1988, Romano1999}, whereas solar-metallicity stars require \teff $\gtrsim 6800$--$6900$ K to avoid depletion \citep{Romano2021}. This $\sim$1100 K threshold shift seems to stem from opacity-driven changes in convective structure: higher metal content (particularly Fe and O) increases opacity, deepening convection zones and enhancing Li destruction \citep[][Sect. 3.5 and Table 3]{Piau2002}. The result resolves the apparent discrepancy between metal-poor halo stars and metal-rich disc populations, demonstrating how metallicity modulates mixing efficiency across Galactic populations.

The absence of detectable star-to-star dispersion in the A(Li)s of warm ($T_{\rm eff} \gtrsim 5700$ K), metal-poor ($-3 \leq$ [Fe/H] $\leq -1.5$ dex) halo dwarfs has long been interpreted as evidence that those stars display the primordial \ali\ \citep[$\simeq$~2.1--2.3; e.g.][]{Rebolo1988, Romano1999}, a conclusion supported also by the measurement of \ali\ $\simeq 2.2$ in the line of sight to Sk~143, an O-type star in the Small Magellanic Cloud \citep{Molaro2024}. However, the existence of the Li meltdown below [Fe/H] $\simeq -3$ dex \citep{Sbordone2010} and the presence of a meagre group of mildly metal-poor halo stars clustering around \ali\ $\simeq 2.7$ in GALactic Archaeology with HERMES data \citep[GALAH;][]{Gao2020} underscore the complex interplay between Li survival and stellar parameters (e.g. effective temperature, metallicity, age).

Additionally, rotation-induced mixing is another relevant factor: models by \citet{Lagarde2012} show that a solar-metallicity 1 $M_{\odot}$ star with moderate initial rotation (23\% of critical) can deplete Li entirely by 4 Gyr, while a non-rotating counterpart retains its initial abundance. Other models also explore the role of rotation-induced mixing, as well as other stellar properties, in order to estimate the variation of Li in stellar interiors and photospheres \citep[e.g.][]{Deal2021}. However, we also refer the interested reader to \citet{LlorenteDeAndres2021} for a somewhat contrasting discussion on how rotation influences \ali. Therefore, Li depletion is indeed a multifactorial process.

Several studies have identified signs of Li depletion correlated with increasing metallicity in dwarf stars with super-solar metallicities, i.e. \feh>0 \citep[e.g.][]{Guiglion2016, Bensby2018, Fu2018, Grisoni2019, Bensby2020}. \citet{Guiglion2019} studied a set of old dwarf stars currently inhabiting the solar vicinity with super-solar metallicities and \ali\ below expectations \citep[by comparing with predictions from the models presented by][that include prescriptions for Li synthesis in novae outbursts and cosmic ray spallation processes following \citealt{Romano1999}]{Chiappini2009}. Indeed, one would expect metal-rich dwarfs to be formed in regions of complex chemical enrichment histories, such as the inner Galaxy \citep[see also][]{Magrini2009, Romano2021}. Hence, \citet{Guiglion2019} raised the hypothesis that this phenomenon could be attributed to a combined effect of stellar evolution and radial migration, the latter being responsible for transporting old metal-rich stars from the inner Galaxy to the solar region.

In a previous paper, \citet{Dantas2023}, we investigated the behaviour of a set of super-metal-rich stars in the solar neighbourhood that most likely had migrated from the inner regions of the Milky Way (MW); in a follow-up letter, we showed that the stars of our sample with \feh>0 seem largely Li-depleted \citep[][]{Dantas2022}, confirming that the Li depletion observed in metal-rich stars is likely caused by an interplay between stellar evolution and radial migration. We also argued that the \ali\ in the atmospheres of warm dwarfs (\teff $\lesssim$ 6800-6900 K) should not be used as a proxy for the \ali\ in the interstellar medium (ISM), in agreement with \citet[][]{Romano2021}. It is worth mentioning that the recent investigation by \citet{LlorenteDeAndres2024} has challenged that stellar motion induces Li depletion.

Following the work established in \citet{Dantas2022, Dantas2023}, we developed a generalised additive model \citep[GAM;][]{HastieTibshirani} to extend the chemical evolution models computed by \citet{Magrini2009} and use them to estimate the birth Galactocentric radii ($R_{\rm b}$) of all the thin disc stars in our sample with varying metallicities, ranging from metal-poor to super-metal-rich \citep[full analysis in][]{Dantas2025}. Therefore, by obtaining both the expected birth radii, $R_{\rm b}$, as well as the current guiding radii, $R_{\rm g}$, we analysed the motion of the stars in our sample, classifying them as either having been relocated from their original orbital radii (i.e. churned inwards or outwards; this motion is commonly known as radial migration), or having remained approximately at their original orbital radii (i.e. blurred or undisturbed stars).

In this paper, we use the results from \citet{Dantas2025} to better understand the \ali\ of our entire sample while evaluating their stellar motions as well as other critical stellar parameters such as effective temperatures, masses, and ages. The core goals are to (i) evaluate whether the correlation between Li depletion and radial migration entails causation, and (ii) to quantify the relative importance of stellar parameters versus migration history through survival analysis. While we analyse how Li depletion relates to stellar parameters \citep[\teff, \feh, \age, as already observed in several earlier studies, such as][]{Deliyannis1990, Pinsonneault1997}, we do not model depletion mechanisms (e.g. rotation-induced mixing). Instead, we focus on testing whether radial migration itself drives depletion—a hypothesis we ultimately disfavour (see Sect. \ref{sec:analysis_results}). Our survival modelling approach complements this by rigorously assessing how \teff, \feh, \age, and motion direction jointly shape Li survival probabilities. For our sample of stars with [Fe/H] $\gtrsim$ -1.0, we adopted $T_{\rm eff} \gtrsim 6800$--$6900$ K as a conservative threshold for identifying objects likely to retain the original Li \citep[see e.g.][]{Gao2020, Romano2021}, although the \teff\ range of our sample is below this threshold. In this work, we adopted the solar Li abundance as $\rm{\ali_{\odot}} = 0.96 \pm 0.05$ dex \citep{Wang2021} and the solar effective temperature as $T_{\rm{eff, \odot}} = 5773 \pm 16$ K \citep{Asplund2021}.

%-------------------------------------------------------------------
\section{Data and methodology} \label{sec:data_method}

We used a sample of 1188 thin-disc dwarf stars from the final data release of the \textit{Gaia}-ESO survey \citep{Gilmore2012, Gilmore2022, Randich2013, Randich2022} observed by the Ultraviolet and Visual Echelle Spectrograph \citep[UVES,][]{Dekker2000}, an instrument of ESO's Very Large Telescope. The atmospheric parameters were derived according to the prescription of \citet{Smiljanic2014} with updates detailed in \citet{Worley2024} and homogenised as described in \citet{Hourihane2023}. For age estimations, we used \textsc{unidam} \citep[][]{Mints2017, Mints2018}, a code that uses PARSEC isochrones \citep[][]{Bressan2012} to estimate stellar ages performing a Bayesian fit using photometric and spectroscopic data from each star. Typical errors are usually around 1 to 2 Gyr \citep[see][]{Mints2017, Mints2018}. For the purposes of this study, it is worth recalling that fast-rotators were removed, meaning we did not account for stars with v$\sin{i} > 10 \rm{\, km\, s^{-1}}$. This dataset is the same as previously described and analysed in \citet{Dantas2023}, and more thoroughly investigated and filtered in \citet{Dantas2025}; for the current study, we summarise the data description into its most particular and relevant features.

As detailed in \citet{Dantas2023}, we applied quality cuts to ensure the overall reliability of the sample, retaining only stars with \texttt{RUWE} $< 1.4$, \texttt{ipd\_frac\_multi\_peak} $\leq 2$, and \texttt{ipd\_gof\_harmonic\_amplitude} $< 0.1$, following \citet[][catalogue validation for \emph{Gaia} early data release 3, \emph{Gaia} EDR3]{Fabricius2021}. The combination of these flags and thresholds is considered robust for removing contamination by binaries, which could include blue stragglers or blue-stragglers-to-be-objects \citep[see][for the latter; but see also \citealt{Kervella2022} on the identification of binaries through \emph{Gaia} EDR3]{Ryan2001}. Such binaries could otherwise bias abundances through mass transfer or merger-related Li depletion. We also excluded \texttt{PECULI}-flagged stars \citep[see][for the flag definition]{Hourihane2023} from the \emph{Gaia}-ESO Survey (i.e. removed those with values other than \texttt{NaN}), which include binaries and other peculiar objects.

Among our 1188 stars, 7 lack Li measurements, resulting in a sample of 1181 stars with either detected Li or upper limits. Li estimations for stars observed by \textit{Gaia}-ESO are described by \citet{Franciosini2022}. In this study, we applied the 3D non-local thermodynamic equilibrium (NLTE) corrections for \ali\ provided by \citet{Wang2021}. Out of our 1181 stars, 13 had lower estimated \ali\ than the grids used in \citet{Wang2021}; we chose to keep these corrections anyway and believe potential errors in the overall analysis are small since they affect $\sim$ 1.1\% out of the 1181 stars.

The stars were classified into six metallicity-stratified groups by making use of hierarchical clustering \citep[HC;][]{Murtagh&Contreras2012, Murtagh2014}, which used as input 21 abundances of 18 individual species, as described in \citeauthor{Dantas2022} (\citeyear{Dantas2022, Dantas2023}) and very clearly shown in \citet[][Fig. 1]{Dantas2025}. It should be noted that \ion{Li}{I} was not one of the abundances used in the HC. Additionally, we kept the original HC numbering adopted throughout our previous works for consistency. The number of stars in each group and their respective parameters are reported in Table \ref{tab:general_properties}, which will be discussed later in the current paper.

In \citet{Dantas2022}, we analysed a subset of stars selected from our main sample, focusing only on those with the six highest \ali\ values and super-solar \feh. The focus on stars with the highest \ali\ values in that work was motivated by the goal of using stars representative of the ISM abundance. However, this is not the case in the current study. Here, we explored the differences and similarities among groups of stars stratified by both metallicity and motion, without limiting the analysis to stars with high \ali\ values. Therefore, in the current study, we examined the entire sample, encompassing stars with a wide range of metallicities, from metal-poor to super-metal-rich.

The detailed description and justification of the survival analysis model are deferred to Sect. \ref{subsubsec:adopted_model}, rather than being introduced in this section. This structure ensures that the model’s parametrisation and assumptions are contextualised alongside later empirical findings, as they are informed by insights and discussions developed throughout this manuscript.

%-------------------------------------------------------------------
\section{Results and discussion} 
\label{sec:analysis_results}

% -------------------------------
\subsection{Confronting stellar parameters with stellar motion}
\label{subsec:param_x_motion}

%-------------
%-------------
\subsubsection{Analysing the complete sample with \ali\ measurements and upper limits}
\label{subsubsec:analysis_limeasurements}

\begin{figure*}
    \centering
    \includegraphics[width=\linewidth]{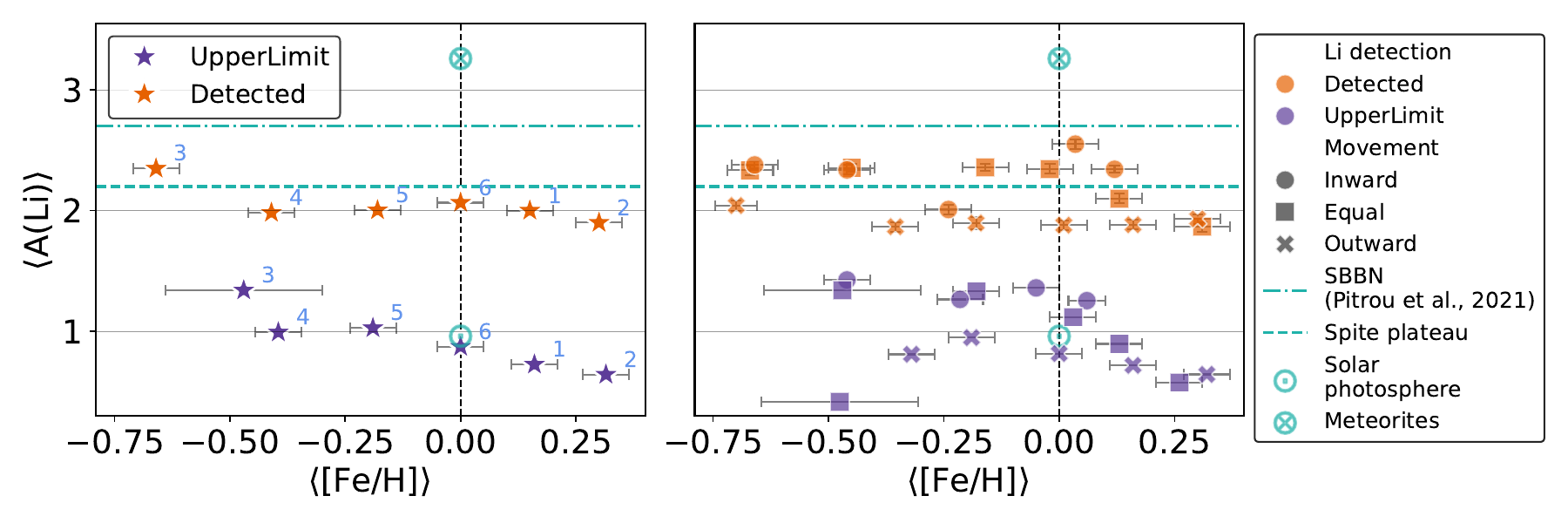}
    \caption{Median lithium abundances ($\langle \rm{\ali} \rangle$) vs median metallicity ($\langle \rm{\feh} \rangle$) with their respective median errors for all the stellar groups in our sample, stratified by HC groups and Li detection (orange markers indicate detected values, and purple markers indicate upper limits). Left panel: Star-shaped markers depict $\langle \rm{\ali} \rangle$ vs $\langle \rm{\feh} \rangle$ for the entire sample, with each marker annotated to indicate the corresponding HC group. Right panel: Similar to the left panel, but further stratified by stellar movement. Circle markers represent stars that moved inwards; x-shaped markers indicate stars that moved outwards; and square markers depict stars with a birth radius similar to their current Galactocentric distance (`Equal'). In cyan, we additionally display the abundances for the Solar photosphere ($\bigodot$), meteorites ($\bigotimes$), the Spite plateau (horizontal dashed line), and a newer SBBN estimate for \ali\ from \citet{Pitrou2021} at 2.7 dex (horizontal dot-dashed line). Note that upper limits do not include error estimations for lithium since no de facto detection was made, as shown in all other plots of the paper.}
    \label{fig:li_feh_all}
\end{figure*}

Figure \ref{fig:li_feh_all} displays two panels with $\langle \rm{\ali} \rangle$ vs $\langle \rm{\feh} \rangle$ for each HC group, the left one shows the entirety of stars in each group and the right panel depicts each group stratified by motion classes\footnote{All parameters used here are presented as median values, derived through previously bootstrapping this sample to estimate uncertainties \citep[see][for further details]{Dantas2023}.}. In both panels, we include the meteoritic value of \ali=3.26 dex \citep{Lodders2009}, which serves as a proxy for the \ali\ of the ISM at the time of the formation of the Solar System; the current solar photospheric abundance [\ali=0.96 dex, \citealt[][]{Wang2021}]; and the Spite plateau as a horizontal dashed line, \ali=2.2 dex \citep[][]{SpiteSpite1982}. However, as the Spite plateau does not agree with theoretical expectations, we also include the standard Big Bang nucleosynthesis (SBBN) prediction of $\simeq$ 2.7 dex \citep[][]{Pitrou2021}, consistent with the 2.6 to 2.8 dex range derived by \citet{Cooke2024} at 95\% confidence. We also note that \citet{Gao2020} identified a group of warm stars with \feh\ between $-$1.0 and $-$0.5 showing a Li plateau at a level consistent with the predictions of SBBN. In addition, \citet{Mucciarelli2022} found a Li plateau in metal-poor giants that, if interpreted taking into account evolutionary mixing processes, is also consistent with SBBN.

In the left panel of Fig. \ref{fig:li_feh_all}, all the HC groups are depicted with star-shaped markers for both detected \ali\ and upper limits (in orange and purple, respectively). The HC group numbers are annotated adjacent to their markers. A mild trend of increasing \ali\ with increasing \feh~up to \feh=0 is observed, after which the trend inverts, except for the most metal-poor group (HC Group 3), which shows the highest $\langle \rm{\ali} \rangle$. The explanation for this higher \ali\ could lie in their overall stellar properties (such as \teff), as we later discuss throughout this paper.

The right panel of Fig. \ref{fig:li_feh_all} further stratifies the HC groups into motion-classified subgroups. Notably, all the stars in all HC groups classified as having moved outwards underwent significant Li depletion, especially evident for those with detected Li (orange x-shaped markers), but also for those with upper limits.

\begin{figure*}
    \centering
    \includegraphics[width=\linewidth]{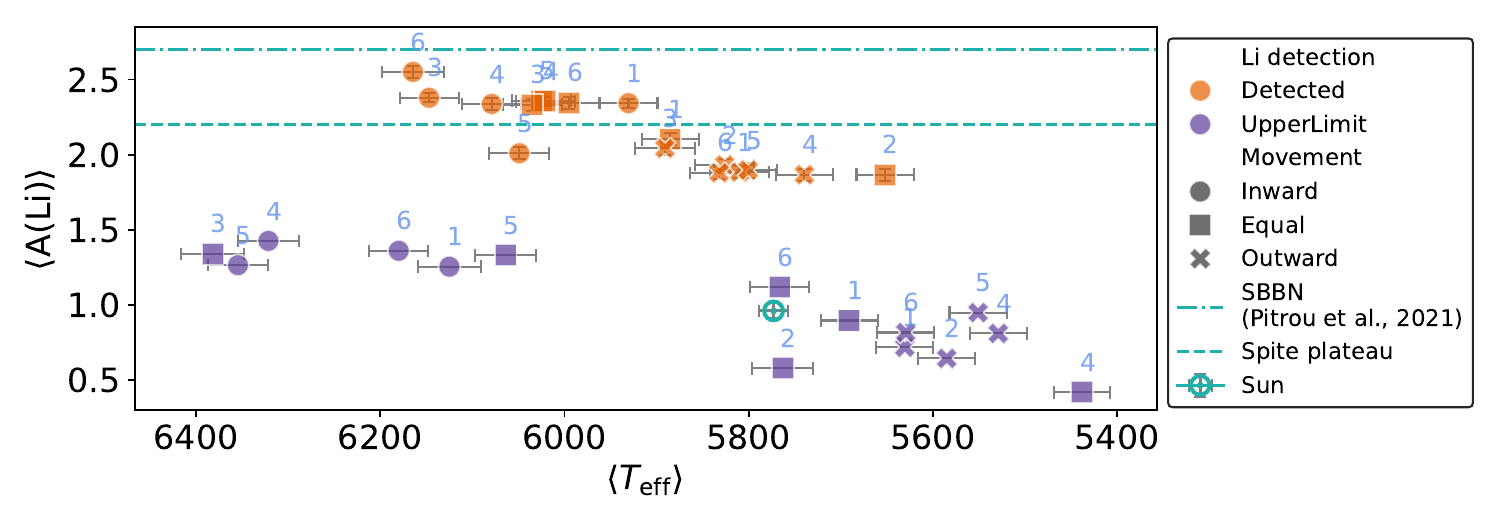}
    \caption{Median lithium abundances ($\langle \rm{\ali} \rangle$) vs median effective temperatures ($\langle T_{\rm{eff}} \rangle$) for all the stellar groups in our sample, stratified by metallicity (through the HC) and Li detection (orange markers indicate detected values, and purple markers indicate upper limits). Circle markers represent stars that moved inwards; X-shaped markers indicate stars that moved outwards; and square markers depict stars with similar birth and current Galactocentric distances (`Equal'). We additionally display the Sun within this parameter space in cyan ($\bigodot$), considering  $T_{\rm{eff, \odot}} = 5773 \pm 16$ K \citep{Asplund2021}; the Spite plateau can be seen through the horizontal dashed cyan line, as well as the SBBN estimate by \citet[][dot-dashed line]{Pitrou2021}.}
    \label{fig:li_teff_all}
\end{figure*}

Figure \ref{fig:li_teff_all} displays $\langle \rm{\ali} \rangle$ vs $\langle T_{\rm{eff}} \rangle$ using the same colours and markers as in the right panel of Fig. \ref{fig:li_feh_all}. The figure clearly shows that different HC groups tend to cluster according to their motion classification, for both detected Li and upper limits. Groups of stars that underwent outwards churning have cooler $\langle T_{\rm{eff}} \rangle$, which largely explains their depletion. In contrast, those that underwent inward churning, with higher $\langle T_{\rm{eff}} \rangle$, tend to preserve their Li. Groups of stars with unchanged motion (i.e. blurred or undisturbed; represented by the square markers) are located between the inward- and outward-churned groups.

Figures \ref{fig:li_feh_all} and \ref{fig:li_teff_all} illustrate the relationship between radial migration and Li depletion: stars that migrated outwards (x-shaped markers) are the coolest in the sample and experienced the most significant Li depletion. Many studies report greater Li depletion in stars with super-solar metallicities \citep[e.g.][]{Guiglion2016, Guiglion2019, Bensby2018, Bensby2020, Stonkute2020}, as these stars predominantly move outwards. In other words, the proportion of metal-rich stars moving outwards is higher than that of metal-poor stars, which affects the overall analysis. By classifying the stars based on different motion classifications, these relationships become evident.

We reinforce that, while complete removal of blue straggler contamination cannot be guaranteed, its impact on our sample seems negligible due to the key observational constraints we described in Sect. \ref{sec:data_method}. Therefore, the majority of the Li-depleted stars in our sample occupy a parameter space distinct from significant blue straggler or blue-straggler-to-be contaminants, which could cause Li depletion due to stellar mergers or mass-transfer.

Our results agree with those of \citet{Zhang2023}, who demonstrate that stars migrating outwards show marked Li depletion in the super-solar \feh\ regime. However, we extend this analysis by exploring additional stellar parameters, such as $T_{\rm{eff}}$, which further contextualise the connection between radial migration and Li depletion, as discussed throughout the current section. Additionally, our results also agree with the findings of \citet{Sun2025}, who found that the most Li-rich stars in their sample (comprised of main-sequence turn-off, MSTO, stars) were formed in the outer disc and migrated inwards.

\begin{figure*}
    \centering
    \includegraphics[width=\linewidth]{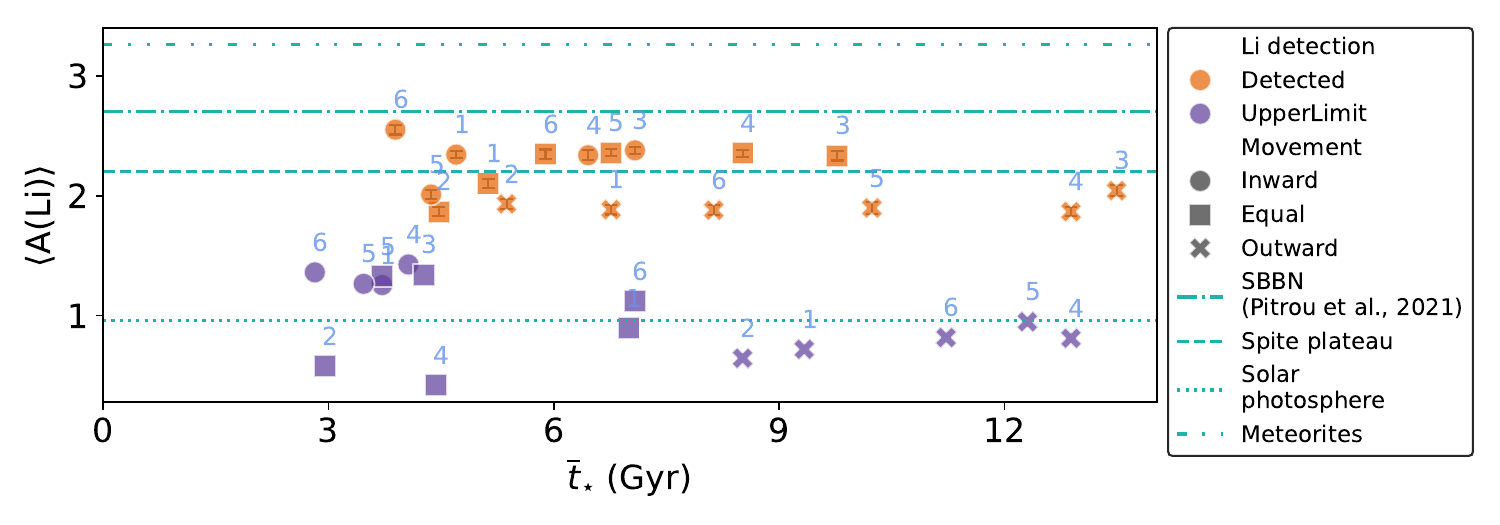}
    \caption{$\langle \rm{\ali} \rangle$ vs \age\ for all the stellar groups of our sample with Li measurements. We depict the solar photosphere and meteorite Li abundances in the shape of dotted and dot-dot-dashed cyan lines, respectively. The HC numbers are annotated adjacent to each respective marker.}
    \label{fig:li_age}
\end{figure*}

Table \ref{tab:general_properties} presents several parameters for the stars in our sample, categorised by HC group numbers, motion classification, and Li measurements. This table extends Table 3 in \citet{Dantas2025}. It is evident that, within each HC group, the stars that underwent Li depletion are the coolest, oldest, and least massive. To better illustrate the relation between $\langle \rm{\ali} \rangle$ and age, \age, we provide an additional figure (Fig. \ref{fig:li_age}).

To further analyse the effects of age and \ali\ we display in Fig. \ref{fig:li_age} $\langle \rm{\ali} \rangle$ vs \age. Distinct markers are placed to distinguish between the different motion classes among the subgroups and different colours to depict the type of Li measurement, similar to Figs. \ref{fig:li_feh_all} and \ref{fig:li_teff_all}; yet, differently from the previous figures, the solar photosphere and meteorite A(Li)s are displayed in the shape of dot-dashed and dotted cyan lines, respectively. The HC group numbers are annotated adjacent to each marker. It is noticeable that within the same groups, stars churning outwards are systematically the oldest, whereas those churning inwards are the youngest, and those that kept their orbital birth radii (blurred or undisturbed) have systematically intermediate ages.

\begin{figure*}
    \centering
    \includegraphics[width=\linewidth]{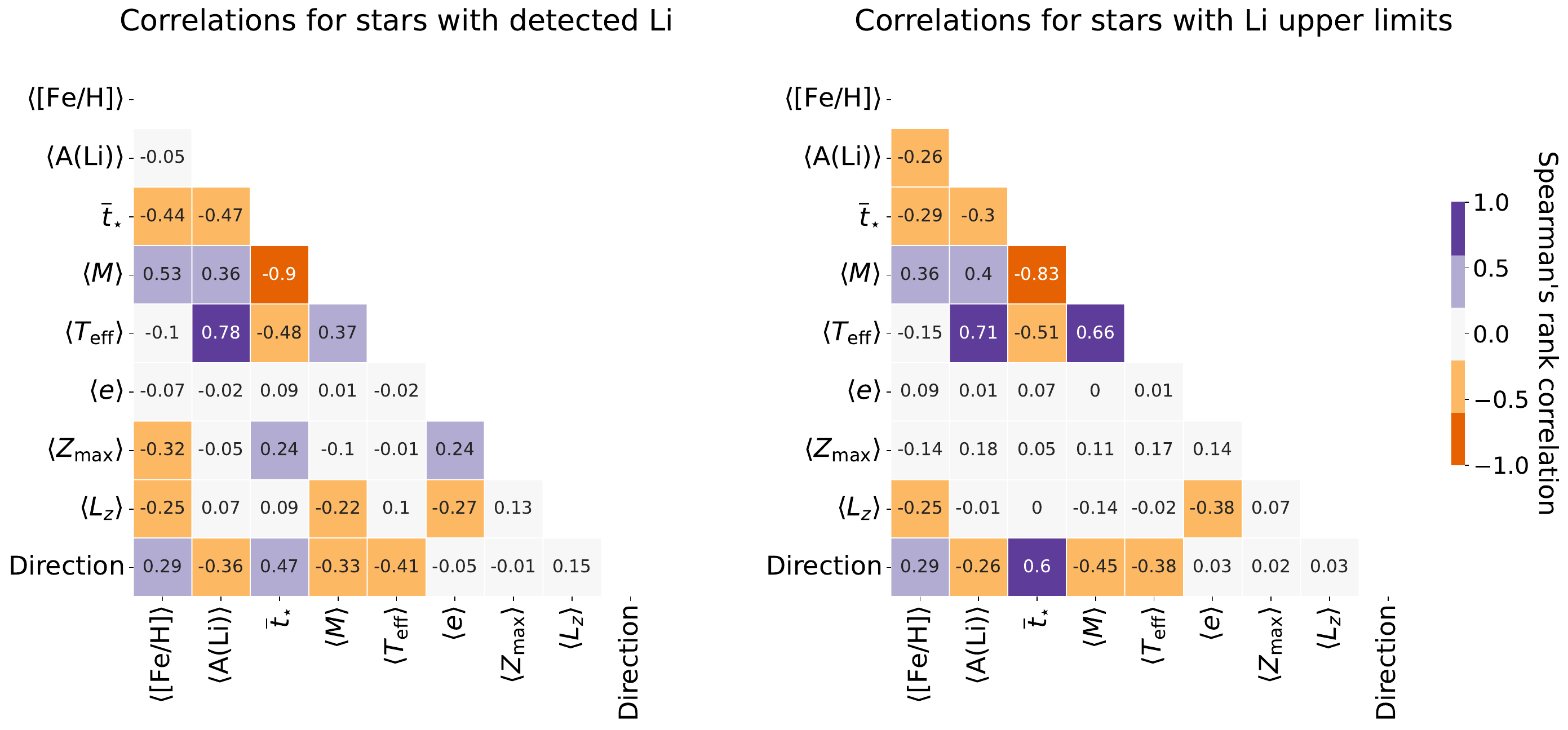}
    \caption{Heat maps displaying the correlations between several parameters for all the stars in the sample stratified by Li detection (detected and upper limits, respectively, from the left to the right). We display the values of $\langle \rm{\feh} \rangle$, $\langle \rm{\ali} \rangle$, \age, $\langle M \rangle$, $\langle T_{\rm eff} \rangle$, \eccentricity, \zmax, \lz, and direction. We considered direction as a numerical value to classify stars churned either outwards (1) or inwards (-1), and either blurred or undisturbed (0).}
    \label{fig:li_correlations}
\end{figure*}

To verify these correlations, we provide a heat map in Fig. \ref{fig:li_correlations}, showing these parameters stratified into two panels by Li detection. These correlations are estimated via Spearman's rank \citep[$\rho$,][]{Spearman1904}. To gauge these correlations, we assigned arbitrary numerical values to the movement directions: churned outwards (1), churned inwards (-1), or equal motion (0).

\renewcommand{\arraystretch}{1.2}

\begin{table*}
    \caption{Overall parameters for all the 1180 stars of our sample in order of decreasing $\langle \rm{[Fe/H]} \rangle$ according to their HC group (`G').}
    \centering
    %\begin{tabular}{cll S[table-format=3] S[table-format=2.2] S[table-format=1.2] S[table-format=1.2] S[table-format=1.2] S[table-format=1.2] S[table-format=1.2] S[table-format=4.1] S[table-format=1.2] S[table-format=4.2]}
    \begin{tabular}{cllrrrrrrrrrr}
    \toprule
    G & Direction & Li det. & $N_{\star}$ & \% & $\langle \rm{\feh} \rangle$ & $\langle \rm{\ali} \rangle$ & $\,\overline{t}_{\star}$ & \,\eccentricity & \zmax & $\langle T_{\rm eff} \rangle$ & $\langle M \rangle$ & $\langle L_z \rangle$ \\
    & & & & & & & Gyr & & kpc & K & M$_{\odot}$ & \multicolumn{1}{c}{$\text{kpc} \cdot \text{km/s}$}\\
    \midrule
    \midrule

    \multirow{5}{*}{2} & \multirow{2}{*}{Inwards}  & D  &   0 &  0.00 &  N/A  & N/A  &  N/A  &  N/A &  N/A &    N/A &  N/A &     N/A \\
                       &                           & UL &   0 &  0.00 &  N/A  & N/A  &  N/A  &  N/A &  N/A &    N/A &  N/A &     N/A \\
                       & \multirow{2}{*}{Outwards} & D  &  39 & 23.93 &  0.30 & 1.93 &  5.37 & 0.13 & 0.50 & 5826.0 & 1.15 & 1605.99 \\
                       &                           & UL & 107 & 66.05 &  0.32 & 0.64 &  8.51 & 0.16 & 0.66 & 5585.0 & 1.02 & 1660.10 \\
                       & \multirow{2}{*}{Equal}    & D  &   5 &  3.07 &  0.31 & 1.86 &  4.47 & 0.14 & 0.71 & 5652.0 & 1.20 & 1384.03 \\
                       &                           & UL &   9 &  5.52 &  0.26 & 0.58 &  2.95 & 0.20 & 0.66 & 5763.0 & 1.23 & 1424.69 \\

    \midrule

    \multirow{6}{*}{1} & \multirow{2}{*}{Inwards}  & D  &   2 &  0.98 &  0.12 & 2.34 &  4.70 & 0.19 & 0.43 & 5930.5 & 1.26 & 1285.42 \\
                       &                           & UL &   1 &  0.49 &  0.06 & 1.25 &  3.72 & 0.19 & 0.98 & 6125.0 & 1.28 & 1314.92 \\
                       & \multirow{2}{*}{Outwards} & D  &  33 & 16.18 &  0.16 & 1.88 &  6.76 & 0.13 & 0.65 & 5810.0 & 1.08 & 1797.93 \\
                       &                           & UL & 120 & 60.00 &  0.16 & 0.72 &  9.33 & 0.13 & 0.62 & 5630.5 & 0.97 & 1790.64 \\
                       & \multirow{2}{*}{Equal}    & D  &  21 & 10.29 &  0.13 & 2.10 &  5.13 & 0.11 & 0.45 & 5885.0 & 1.13 & 1645.05 \\
                       &                           & UL &  22 & 10.78 &  0.13 & 0.89 &  6.99 & 0.16 & 0.51 & 5691.0 & 0.99 & 1541.62 \\

    \midrule

    \multirow{6}{*}{6} & \multirow{2}{*}{Inwards}  & D  &   8 &  2.07 &  0.04 & 2.55 &  3.89 & 0.19 & 0.55 & 6164.5 & 1.18 & 1402.16 \\
                       &                           & UL &   7 &  1.81 & -0.05 & 1.36 &  2.82 & 0.04 & 0.48 & 6180.0 & 1.36 & 1650.56 \\
                       & \multirow{2}{*}{Outwards} & D  &  96 & 25.67 &  0.01 & 1.88 &  8.13 & 0.11 & 0.74 & 5832.0 & 1.02 & 1895.49 \\
                       &                           & UL & 148 & 39.57 &  0.00 & 0.82 & 11.22 & 0.14 & 0.70 & 5629.5 & 0.90 & 1773.25 \\
                       & \multirow{2}{*}{Equal}    & D  &  75 & 19.34 & -0.02 & 2.35 &  5.89 & 0.12 & 0.57 & 5995.0 & 1.07 & 1733.15 \\
                       &                           & UL &  36 &  9.63 &  0.03 & 1.12 &  7.08 & 0.15 & 0.59 & 5766.5 & 0.97 & 1794.89 \\

    \midrule

    \multirow{6}{*}{5} & \multirow{2}{*}{Inwards}  & D  &  22 &  7.46 & -0.24 & 2.01 &  4.37 & 0.15 & 0.58 & 6049.0 & 1.18 & 1677.96 \\
                       &                           & UL &  16 &  5.42 & -0.21 & 1.26 &  3.47 & 0.10 & 0.60 & 6354.5 & 1.22 & 1716.45 \\
                       & \multirow{2}{*}{Outwards} & D  &  84 & 28.47 & -0.18 & 1.90 & 10.23 & 0.14 & 0.70 & 5800.5 & 0.89 & 1837.87 \\
                       &                           & UL &  79 & 26.78 & -0.19 & 0.95 & 12.30 & 0.15 & 0.71 & 5551.0 & 0.83 & 1837.53 \\
                       & \multirow{2}{*}{Equal}    & D  &  76 & 25.76 & -0.16 & 2.36 &  6.76 & 0.13 & 0.59 & 6025.0 & 1.01 & 1804.60 \\
                       &                           & UL &  17 &  5.76 & -0.18 & 1.33 &  3.72 & 0.12 & 0.76 & 6064.0 & 1.21 & 2035.38 \\

    \midrule

    \multirow{6}{*}{4} & \multirow{2}{*}{Inwards}  & D  &  25 & 18.38 & -0.46 & 2.34 &  6.46 & 0.13 & 0.99 & 6079.0 & 1.04 & 1703.92 \\
                       &                           & UL &  18 & 13.24 & -0.46 & 1.43 &  4.06 & 0.16 & 1.00 & 6321.5 & 1.07 & 1961.06 \\
                       & \multirow{2}{*}{Outwards} & D  &  48 & 35.29 & -0.35 & 1.87 & 12.88 & 0.14 & 0.68 & 5739.5 & 0.84 & 1874.76 \\
                       &                           & UL &  22 & 16.18 & -0.32 & 0.81 & 12.88 & 0.15 & 0.61 & 5529.0 & 0.79 & 1898.86 \\
                       & \multirow{2}{*}{Equal}    & D  &  19 & 13.97 & -0.45 & 2.35 &  8.51 & 0.17 & 1.34 & 6021.0 & 0.96 & 1888.64 \\
                       &                           & UL &   4 &  2.94 & -0.47 & 0.42 &  4.43 & 0.15 & 1.83 & 5438.0 & 1.22 & 1826.53 \\

    \midrule

    \multirow{6}{*}{3} & \multirow{2}{*}{Inwards}  & D  &   9 & 42.86 & -0.66 & 2.38 &  7.08 & 0.13 & 1.25 & 6147.0 & 0.97 & 2029.65 \\
                       &                           & UL &   0 &  0.00 &  N/A  & N/A  &  N/A  &  N/A &  N/A &    N/A &  N/A &     N/A \\
                       & \multirow{2}{*}{Outwards} & D  &   2 &  9.52 & -0.70 & 2.04 & 13.49 & 0.13 & 2.06 & 5891.0 & 0.81 & 1826.74 \\
                       &                           & UL &   0 &  0.00 &  N/A  & N/A  &  N/A  &  N/A &  N/A &    N/A &  N/A &     N/A \\
                       & \multirow{2}{*}{Equal}    & D  &   9 & 42.86 & -0.67 & 2.33 &  9.77 & 0.18 & 1.22 & 6035.0 & 0.90 & 2035.21 \\
                       &                           & UL &   1 &  3.03 & -0.47 & 1.34 &  4.27 & 0.20 & 1.52 & 6382.0 & 1.16 & 2439.96 \\
    
    \midrule

    \multirow{6}{*}{Total} & \multirow{2}{*}{Inwards}  & D  &  66 &  5.59 & -0.24 & 2.34 &  4.70 & 0.15 & 0.58 & 6079.0 & 1.18 & 1677.96 \\
                           &                           & UL &  42 &  3.56 & -0.13 & 1.31 &  3.59 & 0.13 & 0.79 & 6250.8 & 1.25 & 1683.51 \\
                           & \multirow{2}{*}{Outwards} & D  & 302 & 25.59 & -0.08 & 1.89 &  9.18 & 0.13 & 0.69 & 5818.0 & 0.96 & 1832.30 \\
                           &                           & UL & 476 & 40.34 &  0.00 & 0.81 & 11.22 & 0.15 & 0.66 & 5585.0 & 0.90 & 1790.64 \\
                           & \multirow{2}{*}{Equal}    & D  & 205 & 17.37 & -0.09 & 2.34 &  6.32 & 0.13 & 0.65 & 6008.0 & 1.04 & 1768.88 \\
                           &                           & UL &  89 &  7.54 & -0.07 & 1.01 &  4.35 & 0.16 & 0.71 & 5764.8 & 1.19 & 1810.71 \\
    
    \bottomrule
    \end{tabular}
    \label{tab:general_properties}
    \tablefoot{Similar table to Table 2 in \citet{Dantas2025}, but with added abundances of Li, \ali\ to further stratify the type of detection (`D' for detected cases and `UL' for upper limits) as shown in column `Li det'. The statistics presented here include 1180 stars, rather than the full sample of 1181 stars, with either detected Li or upper limits. The excluded star lacked Li detection or upper limit in the original data, requiring manual abundance estimation. While incorporated into our evolutionary analysis (Sect. \ref{subsec:survival_analysis}), we exclude it here to maintain uniform measurement treatment. The exclusion has negligible impact ($\leq$0.1\% on all statistics).}
\end{table*}

\renewcommand{\arraystretch}{1.}

Apart from minor differences between the two panels in Fig. \ref{fig:li_correlations}, which depict the correlations for detected Li and upper limits, respectively, the overall trends among the panels are consistent. Figure \ref{fig:li_correlations} confirms expected strong (anti-)correlations, such as those between stellar mass (in solar masses), \mass, and $\overline{t}_{\rm \star}$; as well as between $\langle T_{\rm eff} \rangle$ and $\langle \rm{\ali} \rangle$. We also find a significant correlation between $\overline{t}_{\rm \star}$ and the direction of motion, especially for stars with upper limits for \ali\ (also observable in Table \ref{tab:general_properties}). It is also worth mentioning that no significant (anti-)correlation has been detected regarding eccentricity, \eccentricity, with most of the other parameters, except for angular momentum in the $z$-direction, \lz, which indeed there is a mild correlation; the maximum Galactic height parameter (\zmax) shows weak anti-correlation with \feh\ for stars with detected Li.

Our analysis reveals only a mild anti-correlation between \age\ and \feh, contrary to the stronger trends typically found in chemically homogeneous samples. This results from our sample's composition: thin-disc dwarfs spanning a wide metallicity range ($-1.0 \lessapprox$ \feh $\lessapprox +0.5$), originating from distinct Galactic regions with varying star formation histories. As established in Paper I \citep[specifically refer the reader to Fig. 12 in][]{Dantas2025}, the metal-rich population primarily formed in the inner Galaxy, where intense star formation efficiently enriched the ISM. In contrast, metal-poor stars trace the outskirts, where gradual chemical evolution produces weaker age-metallicity coupling. This intrinsic diversity (encoded in the stars' birth radii) naturally suppresses the global correlation while maintaining coherent sub-population trends.

It is noteworthy that $\langle \rm{\ali} \rangle$ and $\langle \rm{\feh} \rangle$ show no correlation or anti-correlation ($\rho=-0.05$) for stars with detected Li, but a weak anti-correlation for those with upper limits ($\rho=-0.26$). This seems to be an effect caused by the choice of using the full sample instead of the stars with the highest \ali, as previously discussed.

We conclude our discussion of the heat map in Fig. \ref{fig:li_correlations} by verifying some of the results from \citet{Dantas2025} not directly tied to \ali. We observe a mild anti-correlation between \lz\ and \eccentricity, potentially due to the effects of different stellar motion (churn, blur/lack of disturbance), though this should be treated cautiously as no significant (anti-)correlations are seen between the direction of the movement and \eccentricity, \zmax, or \lz.

%------------------------------
\subsubsection{Analysing the stars with missing \ali}
\label{subsubsec:appendix_missing_li}

Beyond our main sample of 1181 stars, we also analysed the 7 stars with missing \ali. The goal here was to assess whether these stars simply had issues with their Li measurements or if their properties could be consistent with an even stronger Li depletion, which would have potentially impaired their measurements or upper limit estimations. In this case, we display their individual properties in Table \ref{tab:li_nan}.

Table \ref{tab:li_nan} shows that all the stars are in the cooler range of the sample, having $T_{\rm eff} \leq 5839$ K, which is way below 6800-6900 K, the estimated threshold of unmodified Li for their high metallicity \citep[][see Sect. 4.1 therein]{Romano2021}. Five of the seven stars show outward radial migration, while only two remain near their birth radii.

\begin{table*}
    \centering
    \caption{Parameters for the remaining seven stars with missing Li in order of decreasing \feh.}
    % \begin{tabular}{ccl S[table-format=1.2] S[table-format=2.2] S[table-format=1.2] S[table-format=1.2] S[table-format=4.1] S[table-format=1.2] S[table-format=4.2]}
    \begin{tabular}{cclcrccccc}
    \toprule
    CNAME & Group & Direction & $\rm{\feh}$ & $\overline{t}_{\star}$ & ~\eccentricity & ~\zmax & $T_{\rm eff}$ & $\langle M \rangle$ & $\langle L_z \rangle$ \\
    identifier & & & & Gyr & & kpc & K & M$_{\odot}$ & \multicolumn{1}{c}{$\text{kpc} \cdot \text{km/s}$}\\
    \midrule
    \midrule
    23101602-0503550 & 2 & Outwards &  0.42 & 11.75 & 0.06 & 0.54 & 5257.0 & 0.91 & 1858.21 \\
    15231484-4208158 & 2 & Outwards &  0.37 &  7.41 & 0.13 & 0.43 & 5691.0 & 1.08 & 1945.49 \\
    10140017-4054309 & 1 & Equal    &  0.13 &  7.76 & 0.05 & 0.25 & 5100.0 & 0.82 & 1806.21 \\
    03395345+0010538 & 6 & Outwards &  0.03 & 10.23 & 0.06 & 0.65 & 5665.0 & 0.92 & 2115.71 \\
    23520084-4303292 & 6 & Outwards &  0.02 &  8.51 & 0.19 & 0.96 & 5839.0 & 0.99 & 1871.57 \\
    03480878-3509232 & 6 & Outwards & -0.04 &  9.23 & 0.07 & 0.52 & 5061.0 & 0.78 & 1854.74 \\
    13300754-4356420 & 5 & Equal    & -0.34 & 11.22 & 0.07 & 0.23 & 5367.0 & 0.82 & 2234.20 \\
    \bottomrule
    \end{tabular}
    \label{tab:li_nan}
    \tablefoot{Data for each of the seven stars without any Li measurements. In this table, the column `G' references the HC group named `Group' in Table \ref{tab:general_properties}. It is worth noting that these are in general very cool stars, with the hottest very close to the unmodified Li threshold, with 5839 K.}
\end{table*}

%------------------------------
\subsection{Survival analysis}
\label{subsec:survival_analysis}

%-------------
%-------------
\subsubsection{Overview of the problem and model reasoning}
\label{subsubsec:overview_modelling}

Selecting an appropriate model to analyse and interpret Li depletion is a challenging endeavour, as it involves a series of assumptions that must be carefully considered in order to reconcile astrophysical expectations with suitable statistical methods. Borrowed from medical research \citep[see, for instance,][]{George2014, Kartsonaki2016}, survival analysis provides a powerful framework for addressing problems involving censored data in Astrophysics, as discussed in \citet[][see their Chapter 10]{FeigelsonBabu2012}. Censored data arise when the true value of a variable is only partially known --- for example, it may be constrained within a range or only known to exceed (or fall below) a given threshold. This situation is common in observational studies, where measurement limitations or design constraints prevent full access to the quantity of interest.

In the context of our study, we encounter three types of Li information: actual measurements, upper limits, and missing values. The latter are likely due to strong depletion, although other factors may be involved, such as when the pipeline was unable to determine Li abundance due to unforeseen issues (e.g. data reduction problems). Statistically, upper limits are typically treated as left-censored observations.

However, our case is somewhat atypical. As established earlier, all stars in our sample likely experienced some level of Li depletion. This is evidenced by both their measured abundances --- none approaching the primordial value from SBBN, estimated by \citet{Pitrou2021} at $\sim$2.6--2.8 dex --- and by their \teff. In an idealised scenario, stars with $T_{\rm eff} \gtrsim 6800$ K and \ali\ close to 2.6 dex could be considered undepleted. Consequently, from a survival analysis perspective, the event of Li depletion has already occurred for all stars in the dataset, making them, in principle, not censored.

Conversely, stars with \ali\ $\gtrsim$ 2.6 dex would be right-censored, as the event (significant depletion below the primordial value) has not yet taken place: they would still be at `risk'. This represents a conceptual shift from the intuitive logic that above a `threshold' means that the event has occurred. In survival analysis terms, we are modelling the drop below a threshold, so stars above it are considered censored because the event is pending.

Nonetheless, adopting 2.6 dex (or other values above it) as a threshold would result in a sample with no censored data, limiting the applicability of survival techniques. For this reason, we instead use a pragmatic threshold corresponding to the Spite plateau \citep[2.2 dex;][]{SpiteSpite1982}, modelling the drop below this value as the event of interest. This allows us to retain censored cases and better capture the depletion process within the sensitivity range of our data. In the following sections, we delve more deeply into the modelling details and the results.

%-------------
%-------------
\subsubsection{The adopted model}
\label{subsubsec:adopted_model}

Our analysis employs penalised splines within a parametric logistic survival framework to characterise the non-linear relationships between stellar parameters and \ali. The model, implemented using the \textsc{survival} package \citep{TherneauGrambsch2000,Therneau2024} in the \textsc{r} environment, incorporates P-spline terms with \texttt{df = 1.5} (degrees of freedom) and penalty constraints for all predictors: \age, \teff, \feh, and stellar motion direction (encoded as $-1$, $0$, and $+1$ for inward-migrating, non-migrating, and outward-migrating stars, respectively). We explicitly excluded \mass\ due to its high collinearity with other parameters (generalised variance inflation factor GVIF$^{1/(2df)} > 3$ during preliminary testing; see Sect. \ref{subsubsec:survival_results}), which would introduce redundancy without improving model performance. The splines' adaptive flexibility captures non-linear trends while the roughness penalty automatically prevents overfitting in data-sparse regions.

The logistic distribution was selected for its dual capacity to model both extreme \ali\ variations and typical abundance regimes. Its heavier tails accommodate chemically peculiar stars with anomalous Li content, while the sharper central peak provides enhanced precision for the bulk population. Crucially, unlike strictly positive distributions such as Weibull or LogNormal (which are suitable for modelling positive response variables, such as time --- or `survival time'), the logistic framework naturally handles the full dynamical range of observed \ali\ values without artificial truncation.

For the analysis of Li depletion events, the data are classified into four main scenarios. First, for stars with measured \ali\ $>$ 2.2, we consider that a significant depletion event has not yet occurred, so the data are right-censored. In this case, both the upper and lower bounds, $Y_i^U$ and $Y_i^L$, are equal to the observed \ali, reflecting that the star's Li has not yet dropped below the threshold. Second, for stars with measured \ali\ $\leq$ 2.2, the event time is exactly known, as \ali\ has already dropped below the threshold (strong depletion is happening). Therefore, both the upper and lower bounds are set to the observed \ali, and the data are not censored, representing an exact observation. Third, for stars with \ali\ flagged as upper limits, the depletion event has occurred, but the exact timing is uncertain. In this case, the upper bound, $Y_i^U$, is set to the measured upper limit \ali, while the lower bound, $Y_i^L$, is set to $-\infty$, indicating uncertainty about when the depletion occurred. This scenario is considered interval-censored, as the event has occurred but its exact timing is unknown. Lastly, for stars with missing Li data, we assume that the star is depleted, but the timing of the depletion is uncertain. The upper bound, $Y_i^U$, is set to $\texttt{low\_li}$, which is the smallest observed \ali\ in the entire sample minus 0.05 dex, and the lower bound is set to $-\infty$, reflecting complete uncertainty about when the depletion event took place. This is also treated as interval-censored data. We refer the reader to Table \ref{tab:censoring_definitions} for a summary of the censoring definitions and intervals considered in each type of Li status.

To summarise how each data type is handled in the survival analysis, we classify the observations according to their interpretation and censoring role, which we described above. With this approach, we intend to identify, given the Li status of a star, the factors that determine how far along the depletion process a star is. We then proceed to expose the mathematical representation of the adopted model, which is as follows.

\begin{equation}
     Y_i \sim \text{Logistic}(\mu_i, \sigma),
\end{equation}

\noindent where $Y_i$ is the latent survival time [i.e. the log \ali\ at which the threshold crossing occurs], and $\mu_i$ is the linear predictor for the $i$th star, expressed as a sum of non-linear spline terms:

\begin{equation}  
    \mu_i = \beta_0 + f_1(\overline{t}_{{\star},{i}}) + f_2(T_{{\rm eff},i}) + f_3(\text{[Fe/H]}_i) + f_4(\text{direction}_i),  
\end{equation}

\noindent where $\beta_0$ is the model intercept and $f_1, \dots, f_4$ denote penalised smoothing splines (P-splines) with 1.5 degrees of freedom, applied to each predictor. These splines transform the raw covariates into flexible, smooth functions that accommodate non-linear dependencies while maintaining computational stability through a quadratic penalty on the second derivatives. The logistic distribution's parameters --- location ($\mu_i$) and scale ($\sigma$) --- are thus conditioned on the smoothed predictor space, enabling the model to capture complex relationships without a priori assumptions about functional forms. For our application, the survival function for the logistic distribution is given as

\begin{equation}
    S(t) = \frac{1}{1 + \exp\left(\frac{t - \mu_i}{\sigma}\right)},
\end{equation}

\noindent where $t$ represents \ali\ rather than chronological time, and $\mu_i$ is the spline-predicted location parameter for the $i$-th star. Here, $S(t)$ quantifies the probability that a star's \ali\ remains above the threshold $t$ (2.2 dex), effectively framing Li depletion as a `survival' process in abundance space. This approach naturally handles the decline of \ali\ while accounting for covariate-dependent non-linearities through the spline terms $f_1, \dots, f_4$.

\begin{table*}
\caption{Censoring definitions for Li depletion events applied in our survival analysis.}
    \centering
    \begin{tabular}{lccll}
        \toprule
        Condition                     & $Y_i^U$          & $Y_i^L$   & Censoring type       & Description \\ 
        \midrule
        \midrule
        Li is measured and $> 2.2$    & \ali\            & \ali\     & Right-censored       & The event has not yet occurred \\
        Li is measured and $\leq 2.2$ & \ali\            & \ali\     & Not censored (exact) & The event time is known \\
        Li is an upper limit          & \ali\            & $-\infty$ & Interval-censored    & The event has occurred but the time is uncertain \\
        Li is missing                 & \texttt{low\_li} & $-\infty$ & Interval-censored    & We assume Li depletion \\ 
        \bottomrule
    \end{tabular}
    \label{tab:censoring_definitions}
    \tablefoot{The variables $Y_i^L$ and $Y_i^U$ represent the latent times corresponding to the lower and upper bounds of the depletion event.}
\end{table*}

Importantly, survival analysis is not the only option for analysing the relation between \ali\ and other variables. Traditional regression techniques remain well-suited to modelling Li as a continuous variable dependent on parameters such as the ones herein used. Our use of survival methods offers an alternative perspective on the depletion process by explicitly modelling threshold crossing.

%-------------
%-------------
\subsubsection{Survival analysis results and discussion}
\label{subsubsec:survival_results}

The penalised spline regression model achieved exceptional explanatory power for \ali\ variations, with a scale parameter $\sigma = 0.44$ indicating highly precise predictions. In logistic survival models, $\sigma$ represents residual dispersion; values $< 1$ (here 0.44) indicate predictions closely track observations, with $\sigma$ closer to 0 approaching deterministic relationships. The low $\sigma$ suggests \ali\ depletion is highly predictable from these parameters, with minimal unmodelled astrophysical scatter.

The deviance statistics ($\chi^2 = 917.66$ on 7.59 effective degrees of freedom, $p \ll 0.001$)\footnote{We interpret $p$-values cautiously given their well-documented limitations \citep[e.g.][]{Lin2013, Nuzzo2014, Halsey2015, Wasserstein2016}, prioritising effect sizes and information criteria as more reliable evidence.} further underscore the model's explanatory strength. Information criteria \citep[AIC = 2047.57, BIC = 2086.13; AIC and BIC corresponding respectively to Akaike and Bayesian Information Criteria; see, respectively,][]{Akaike1974, Schwarz1978} validated superiority over alternatives (including Gaussian distributions and unpenalised splines), with lower values indicating better parsimony-adjusted fit \citep{BurnhamAnderson2002}. Collinearity diagnostics confirmed no concerning multicollinearity among predictors, with GVIFs all below 3 [GVIF$^{1/(2\rm{df})}$ values: 1.10 (\age), 1.06 (\teff), 1.08 (\feh), 1.09 (direction of motion), well below the conservative threshold of 2.5; see, for instance, \citealt{FoxMonette1992}].

Three key trends emerged from the parameter rankings\footnote{The $z$-score ($z = \text{Coefficient}/\text{Std. Error}$) measures each predictor's relative importance in driving \ali\ variations. Larger absolute values ($|z|$) indicate stronger empirical evidence for the parameter's influence, allowing hierarchical ranking: for example, $|z| = 19.10$ (\teff) implies 3$\times$ greater impact than $|z| = 6.03$ (direction of motion).}:

\begin{enumerate}[(i)]
    \item \teff\ dominated as the most influential factor ($z = 19.10$, $p < 2 \times 10^{-16}$), with positive coefficients (1.08 to 4.24) confirming its protective role: hotter stars retain more Li until saturation above $\sim$6000 K\footnote{These \teff\ results are specific to our stellar sample, which lacks hotter stars (\teff $\gtrsim 6800$ K) that typically preserve Li more efficiently, as discussed earlier (Sect. \ref{sec:intro}).}. The $z$-value's magnitude reflects \teff's overwhelming dominance over other parameters, aligning with its established role in suppressing Li destruction through convective inhibition \citep[e.g.][see their figure 4; but see also \citealt{Piau2002}]{Deal2021}.

    \item \feh\ showed the strongest negative impact ($z = -10.36$, $p < 2 \times 10^{-16}$), where a 0.1 dex metallicity increase accelerates depletion by 15--20\% (derived from $\exp(-0.54 \times 0.1) \approx 0.85$ to $\exp(-4.12 \times 0.1) \approx 0.66$; i.e. 15--34\% faster depletion). We conservatively report 15--20\% for typical \feh\ ranges. The absolute $z$-score magnitude confirms metallicity as the secondary driver of depletion, likely through enhanced opacity and mixing \citep[e.g.][]{Piau2002}.

    \item \age\ exhibited non-linear depletion acceleration after $\sim$2 Gyr ($z = -7.75$, $p = 9 \times 10^{-15}$), with spline coefficients shifting from $-0.15$ to $-2.25$ at this inflection point -- aligning with models of deep convection onset \citep[see e.g.][and empirical evidence discussed in \citealt{Carlos2019, Romano2021}; and \citealt{Martos2023}, to mention a few]{Baraffe2017}.
    
\end{enumerate}  

Motion direction, while considered statistically significant ($z = 6.03$, $p = 1.6 \times 10^{-9}$), had negligible practical impact ($\Delta \rm{\ali} < 0.1$ dex, depicting a near-constant trend). The modest $z$-value relative to intrinsic parameters reinforces that kinematic history is secondary to stellar physics.

\begin{figure*}
    \centering
    \includegraphics[width=\linewidth]{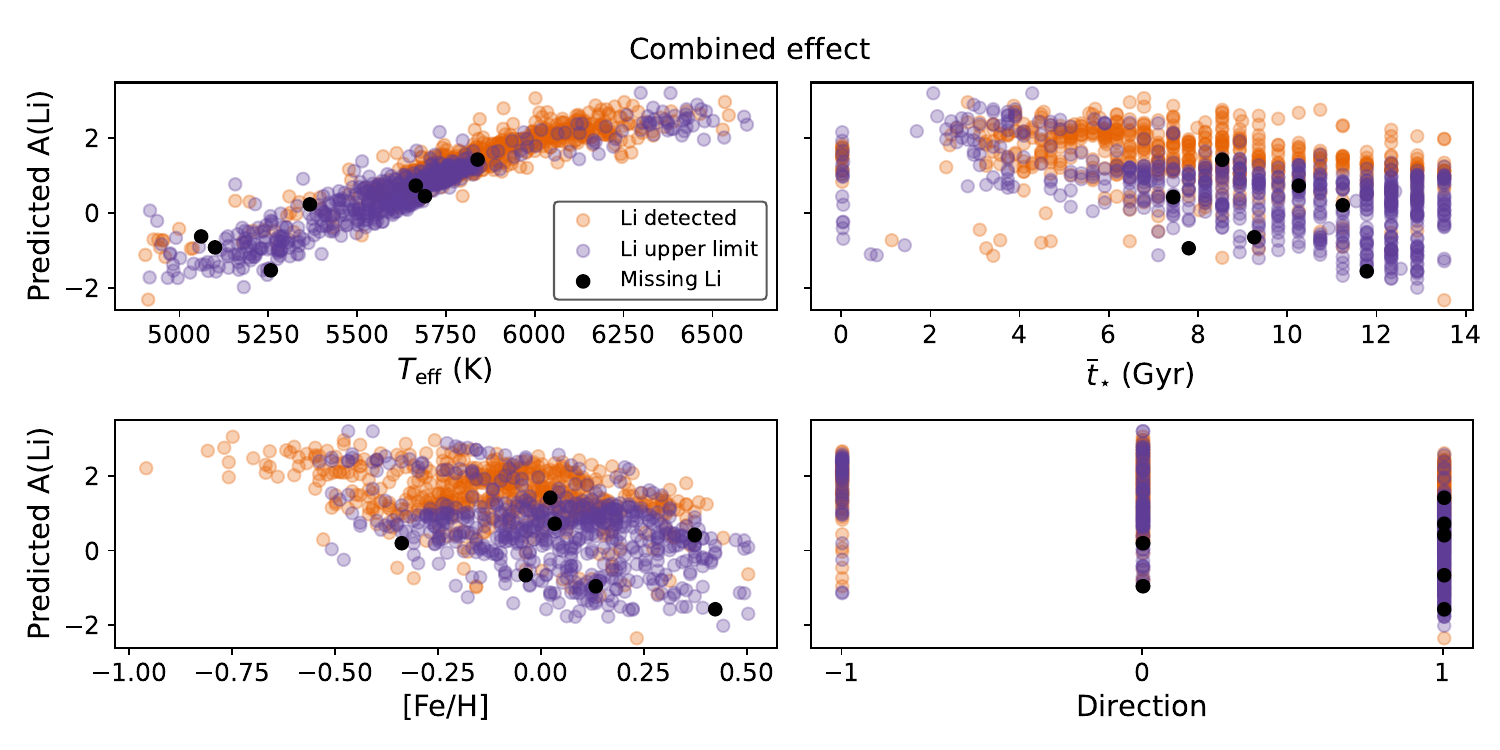}
    \caption{Combined effects of \teff, \age, \feh, and direction of motion on the predicted survival time (in this case, the predicted \ali), accounting for all covariates simultaneously. Predictions are shown for stars with detected \ali\ measurements (orange), upper limits (purple), and missing \ali\ values (black), with colours consistent across all figures. The curves represent the model’s output from the parametric survival regression (\texttt{survreg} with logistic distribution), integrating the contributions of all covariates.}
    \label{fig:li_predictions}
\end{figure*}

\begin{figure*}
    \centering
    \includegraphics[width=\linewidth]{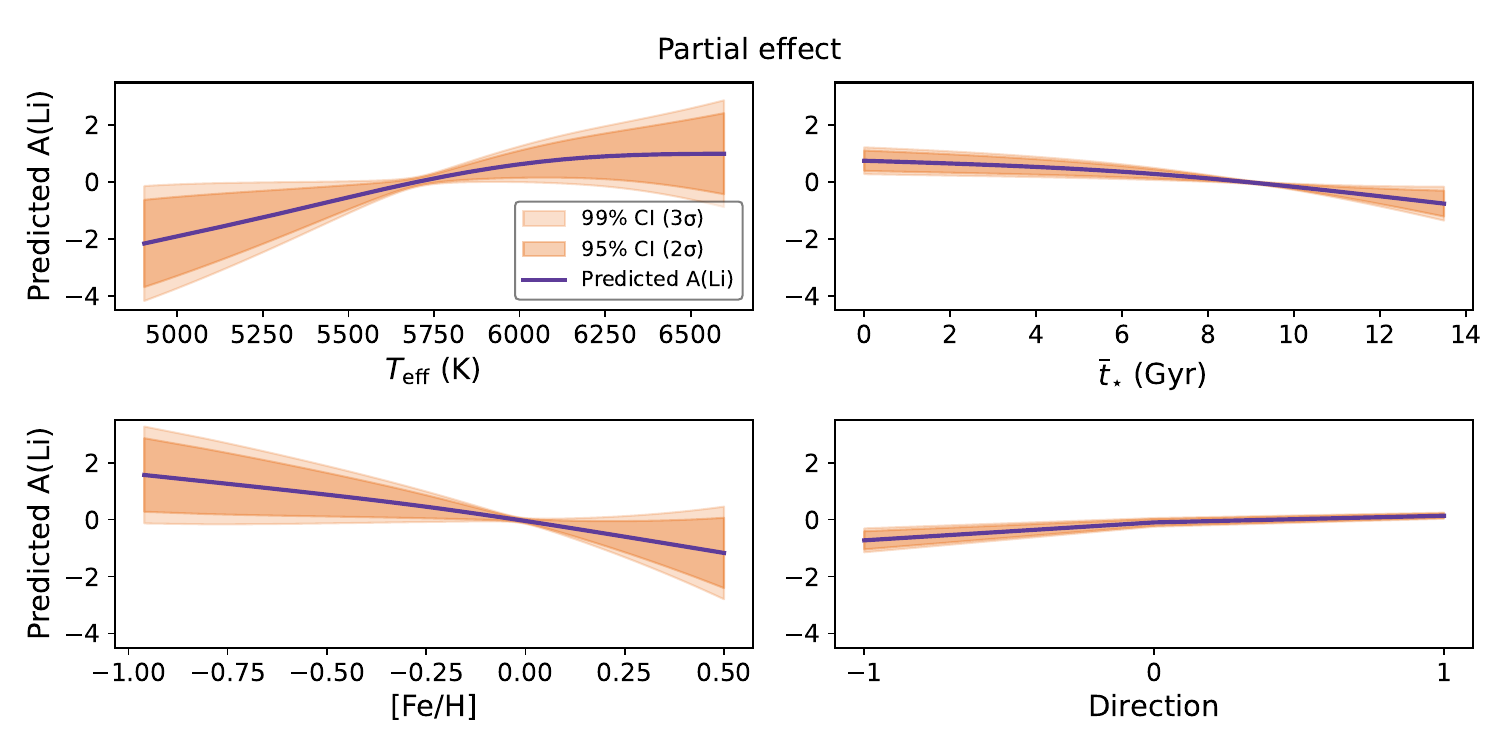}
    \caption{Partial (isolated) effects of \teff, \age, \feh, and direction of motion on the predicted survival time (in this case, the predicted \ali), derived from a parametric survival model (\texttt{survreg} with logistic distribution). Each subplot illustrates the dependence of the predictor on one covariate, while holding all other covariates fixed: \teff, \age, and \feh\ are set to their mean values to represent central tendencies, while the direction of motion is set to zero as a neutral reference point for directional data. Uncertainty in the partial effects is quantified by 95\% ($\sim 2 \sigma$) and 99\% ($\sim 3 \sigma$) confidence intervals, which account for the flexibility of the penalised splines (\texttt{pspline}) used to model each covariate.}
    \label{fig:li_partial}
\end{figure*}

The logistic distribution's heavier tails proved essential for capturing both metal-poor stars with primordial \ali\ and metal-rich outliers, a pattern Gaussian assumptions would underestimate. The results for the full survival analysis can be verified in Fig. \ref{fig:li_predictions}, while the isolated (partial) effects of each parameter to the survival model are depicted in Fig. \ref{fig:li_partial}.

This analysis reveals a coherent astrophysical narrative for \ali\ evolution in stars (quantified through full predictions and partial effects in Figs. \ref{fig:li_predictions} and \ref{fig:li_partial}, respectively): \teff\ governs the primary survival threshold, metallicity seems to control the efficiency of depletion mechanisms, and \age\ determines the onset of destructive mixing processes. Crucially, these intrinsic properties seem to operate hierarchically --- with \teff\ setting the initial survival conditions, \feh\ modulating the depletion rate, and \age\ activating late-stage mixing --- while kinematic history plays a negligible role. The logistic framework successfully unifies these processes, capturing both the bulk behaviour of \ali\ and the extremes at metal-rich and metal-poor regimes. This supports a paradigm where Li depletion is predominantly driven by the physics of stellar interiors, with Galactic dynamics contributing minimal secondary effects (if at all).

\section{Summary and conclusions} \label{sec:conclusions}

In this paper, we used a sample of Galactic thin-disc stars retrieved from the final data release of the \textit{Gaia}-ESO survey. All stars were previously classified into six metallicity-stratified groups via HC, with abundances ranging from super-metal-rich to metal-poor. In \citet{Dantas2025}, we used a GAM to extend the MW's chemical enrichment models described by \citet{Magrini2009} to estimate the probable birth radii of each star in our sample. This allowed us to classify these stars as having churned inwards or outwards (i.e. having suffered radial migration) or as having kept their birth radii (i.e. being either undisturbed or blurred). Using both dynamical classification and parametric survival analysis, we then probed the role of stellar evolution and radial migration in the depletion of Li, as a follow-up to our previous investigation performed in \citet{Dantas2022}. Our survival modelling complemented the dynamical approach by quantifying how \teff, \feh, \age, and direction of motion (migration or lack thereof) jointly influence Li survival probabilities. We successfully identified the relationship between the A(Li)s and the movement of these stars. Our main conclusions are as follows.
\begin{enumerate}
    \item In our sample of FGK-type stars, we observe that those that migrated outwards are typically older, cooler, less massive, and exhibit higher Li depletion when compared to their stellar counterparts that either churned inwards or kept their original orbital radii within each HC group. Most importantly, this result is independent of metallicities (at all ranges). These physical parameters (\teff, \feh, \age) are known to be the classical culprits of Li depletion.

    \item On the other hand, the stars in our sample that migrated inwards are generally younger, hotter, and more massive, whereas those that kept their birth orbital radii have intermediate features (i.e. in between outward- and inward-churned stars). With higher temperatures, these stars manage to more efficiently preserve their photospheric Li, because of their thinner convective layers.

    \item Still, the \ali\ of our stellar sample lies systematically below both the meteoritic value and A(Li)$_{\rm max}$ values of younger \textit{Gaia}-ESO iDR6 open clusters, where undepleted A(Li)s are estimated \citep{Romano2021}. This discrepancy arises because our field stars are predominantly older than 1-2 Gyr, whereas the \textit{Gaia}-ESO clusters with undepleted Li either contain younger stars ($<$1 Gyr) or host stars hot enough to preserve their original Li \citep[see][their Table 2]{Romano2021}. This is strong evidence that all the stars in the current sample suffered Li depletion to some extent.

    \item We provide an assessment of the seven stars with missing Li measurements. Their main features are consistent with a drastic Li depletion, which would explain why it has not been measured in their atmospheres. These stars are also on the cooler side of the sample, with \teff $\leq 5839$ K, which is way below the unmodified Li threshold for their metallicity \citep[$\sim$ 6800-6900 K,][]{Romano2021}.

    \item Our survival analysis suggests that \ali\ variations across the stellar population may be shaped by three interconnected physical regimes: a temperature-dependent survival threshold, metallicity-driven depletion efficiency, and age-triggered mixing. This tripartite structure tentatively explains both the bulk of the \ali\ distribution and its outliers: from metal-poor stars showing slightly enhanced Li preservation to metal-rich stars exhibiting more rapid depletion. While kinematics appear to play a negligible role in our models, implying predominantly local stellar processes, we caution that minor residual uncertainties (such as small imperfections in NLTE corrections or potential past rotational history effects) could contribute to residual scatter. These findings broadly align with theoretical frameworks of Li evolution in stars \citep[e.g.][]{SomersPinsonneault2015}, though we emphasise that our statistical approach complements rather than supersedes existing physical models.

    \item Building on the discussion in \citet{Guiglion2019}, \citet{Romano2021}, and \citet{Dantas2022}, we emphasise that the Li abundance in stars displaying characteristics indicative of potential Li depletion (such as those presented in this study, e.g. $T_{\rm eff} \lesssim 6800$ K) should not be regarded as a reliable indicator of the ISM Li abundance.

\end{enumerate}

The high frequency of Li depletion observed in stars with super-solar metallicities \citep[as shown in earlier studies, such as][]{DelgadoMena2015, Bensby2018, Guiglion2019, Dantas2022} emerges naturally from our analysis, especially through the survival regression statistics, as a consequence of three hierarchical stellar properties: \teff\ (setting the survival threshold), \feh\ (modulating depletion efficiency), and \age\ (activating late-stage mixing). These parameters dominate the observed depletion patterns, particularly in the solar vicinity where metal-rich, outward-migrating stars are overrepresented compared to their metal-poor, inward-migrating counterparts \citep[as shown in Paper I;][]{Dantas2025}.

Our analysis, especially our logistic survival model, demonstrates that the apparent metallicity dependence arises through two complementary mechanisms: (i) the covariance of \feh\ with the key depletion drivers (\teff\ and \age), reflecting a selection effect in our sample \citep[likely tied to \textit{Gaia}-ESO's target selection;][]{Stonkute2016}; and (ii) the enhanced radiative opacity in metal-rich stars that facilitates Li destruction through deeper convective envelopes. Together, this dual interpretation bridges our statistical results with observational constraints and stellar theory, while agreeing with conclusions from previous studies \citep[e.g.][]{Randich2020, Charbonnel2021}.

Therefore, the connection between radial migration and Li depletion appears secondary to these intrinsic stellar properties. We theorise that outward churning operates on timescales comparable to stellar evolution, causing migrated stars to exhibit both old ages and characteristic \teff\ where Li destruction becomes efficient. This explains why the Sun \citep[which most likely migrated from inner Galactocentric distances; see e.g.][]{Tsujimoto2020, Dantas2025} follows the same depletion trend as other cool old metal-rich stars. The survival framework quantitatively supports the following picture: while kinematic history was included as a potential predictor, its negligible contribution in our spline-based model confirms that migration primarily correlates with depletion through its association with temperature, metallicity, and age (the dominant parameters identified in our hierarchical analysis), rather than acting as an independent physical mechanism.

Critically, our analysis demonstrates that the observed correlation between stellar motion and Li depletion does not imply causation: a key distinction, operating through well-characterised stellar physics distinction, that is highlighted throughout this work and supported by the hierarchical dominance of \teff, \feh, and \age\ in our survival model.

%-------------------------------------------------------------------
\section*{Data availability}

The stellar catalogue and chemo-dynamic parameters used in this work are publicly available through the VizieR service \citep{vizier} at the Centre de Données astronomiques de Strasbourg (CDS). The specific dataset, originally published in Paper I, can be accessed via DOI: \url{https://doi.org/10.26093/cds/vizier.36960205}.

%-------------------------------------------------------------------
\begin{acknowledgements}

The authors thank the referee, Piercarlo Bonifacio, for the kind of suggestions that contributed to the improvement of our manuscript. MLLD acknowledges Agencia Nacional de Investigación y Desarollo (ANID), Chile, Fondecyt Postdoctorado Folio 3240344. MLLD and PBT acknowledge ANID Basal Project FB210003. RS acknowledges support from the National Science Centre, Poland, project 2019/34/E/ST9/00133. GG acknowledges support by Deutsche Forschungs-gemeinschaft (DFG, German Research Foundation) – project-IDs: eBer-22-59652 (GU 2240/1-1 "Galactic Archaeology with Convolutional Neural-Networks: Realising the potential of Gaia and 4MOST"). This project has received funding from the European Research Council (ERC) under the European Union’s Horizon 2020 research and innovation programme (Grant agreement No. 949173). RSS acknowledges the support from the São Paulo Research Foundation under project 2024/05315-4. PBT thanks Fondecyt Regular 2024/1240465. MLLD thanks Miuchinha for her long companionship, love, and support. This work made use of the following on-line platforms: \texttt{slack} (\url{https://slack.com/}), \texttt{github} (\url{https://github.com/}), and \texttt{overleaf} (\url{https://www.overleaf.com/}). This work was made with the use of the following \textsc{python} packages (not previously mentioned): \textsc{matplotlib} \citep{Hunter2007}, \textsc{numpy} \citep{Harris2020}, \textsc{pandas} \citep{mckinney-proc-scipy-2010}, \textsc{seaborn} \citep{Waskom2021}. Based on data products from observations made with ESO Telescopes at the La Silla Paranal Observatory under programme ID 188.B-3002.

\end{acknowledgements}

%-------------------------------------------------------------------
\bibliographystyle{aa}        % style aa.bst
\bibliography{paper}

\end{document}